\documentclass[preprint,11pt]{aastex}

\newcommand{\BS}{BS~16920--017} 
\newcommand{\HD}{HD~4306} 
 
\newcommand{\teff}{$T_{\rm eff}$}
\newcommand{\logg}{$\log g$}
\newcommand{\loggf}{$\log gf$}
\newcommand{\vt}{$v_{\rm micro}$}
\newcommand{\loge}{$\log \epsilon$}

\shorttitle{Abundance analysis of {\BS}}
\shortauthors{Honda et al.}

\begin{document}

\title{Spectroscopic Studies of Extremely Metal-Poor Stars with the
Subaru High Dispersion Spectrograph. V. The
Zn-Enhanced Metal-Poor Star {\BS}\footnote{Based on data
collected at the Subaru Telescope, which is operated by the National
Astronomical Observatory of Japan.}}

\author{Satoshi Honda\altaffilmark{2,3}, Wako Aoki\altaffilmark{4,5},
Timothy C. Beers\altaffilmark{6}, Masahide Takada-Hidai\altaffilmark{7}}

\altaffiltext{2}{Gunma Astronomical Observatory, Takayama-mura, Agatsuma, Gunma 377-0702, Japan}
\altaffiltext{3}{Current address: Kwasan Observatory, Kyoto University, Ohmine-cho Kita Kazan, Yamashina-ku, Kyoto, 607-8471, Japan; honda@kwasan.kyoto-u.ac.jp.}

\altaffiltext{4}{National Astronomical Observatory, Osawa, Mitaka, Tokyo,
181-8588 Japan; email: aoki.wako@nao.ac.jp}

\altaffiltext{5}{Department of Astronomical Science, School of Physical Sciences, The Graduate University of Advanced Studies, Mitaka, Tokyo 181-8588, Japan;}

\altaffiltext{6}{Department of Physics \& Astronomy and JINA: Joint Institute for Nuclear
Astrophysics, Michigan State University, East Lansing, MI 48824-1116;
email: beers@pa.msu.edu}

\altaffiltext{7}{Liberal Arts Education Center, Tokai University,
Hiratsuka, Kanagawa, 259-1292, Japan; email:
hidai@apus.rh.u-tokai.ac.jp}

\begin{abstract} 

We report Zn abundances for 18 very metal-poor stars studied in our previous
work, covering the metallicity range $-3.2 < $ [Fe/H] $ < -2.5$. The [Zn/Fe] values of
most stars show an increasing trend with decreasing [Fe/H] in this metallicity
range, confirming the results found by previous studies. However, the extremely
metal-poor star {\BS} ([Fe/H]$ = -3.2$) exhibits a significantly high [Zn/Fe]
ratio ([Zn/Fe] = +1.0). Comparison of the chemical abundances of this object with
{\HD}, which has similar atmospheric parameters to {\BS}, clearly demonstrates a
deficiency of $\alpha$ elements and neutron-capture elements in this star, along
with enhancements of Mn and Ni, as well as Zn. The association with a hypernova
explosion that has been proposed to explain the high Zn abundance ratios found
in extremely metal-poor stars is a possible explanation, although further studies
are required to fully interpret the abundance pattern of this object.

\end{abstract}
\keywords{nuclear reactions, nucleosynthesis, abundances --- stars:
abundances --- stars: Population II --- supernovae: general ---
stars: individual (BS~16920--017)}

\section{Introduction}\label{sec:intro}

The chemical compositions of very metal-poor stars have been intensively
investigated to provide observational constraints on the yields of early
generations of massive stars and the resulting elemental enrichment in the
Galaxy. In particular, extremely metal-poor stars (EMP; [Fe/H] $ < -3$)\footnote{We use
the usual notation [A/B]$\equiv \rm{log}_{10} (N_A/N_B)_*-\rm{log}_{10}(N_A/N_B)
_\odot$ and log$\epsilon(\rm A) \equiv\rm{log}_{10}(N_A/N_H)+12.0$, for elements
A and B. Also, the term ``metallicity'' will be assumed here to be equivalent to
the stellar [Fe/H] value.} have been argued to reflect the result of individual
nucleosynthesis processes associated with the supernovae explosions 
of early generation stars \citep[e.g.,][]{beers05}.

Zinc, along with Ge, is often considered the heaviest of the Fe-peak elements,
and its production by core-collapse supernovae is expected to be dependent on
details of the explosions. Observational studies in the past decades have
revealed a clear increasing trend of [Zn/Fe] ratios with decreasing metallicity
in the range of [Fe/H]$\lesssim -2.5$ \citep[e.g.,][]{cayrel04}. Such a trend
could not be explained by previously studied core-collapse supernova models.
Thus, large contributions of energetic supernovae, called hypernovae, have been
suggested as a possible source of the higher [Zn/Fe] ratios in the early stages
of chemical enrichment of the Galaxy \citep{umeda02}. Chemical evolution models
including hypernovae yields reproduce, at least qualitatively, the observed trend
of [Zn/Fe] at very low metallicity \citep[e.g.,][]{kobayashi06,tominaga07}.
Other scenarios to explain Zn excesses at extremely low metallicity have been
also discussed. For example, Zn might be also synthesized by neutron-capture
processes during helium burning in the shells of low- to intermediate-mass AGB
stars (main s-process) and in the cores of massive stars (weak s-process) , as
has been investigated by \citet{baraffe92}, \citet{matteucci93}, and
\citet{raiteri93}. However, it remains unclear how effective these contributions
might be at extremely low metallicity.

Another interesting observational result on Zn abundances at very low
metallicity is the rather small scatter of [Zn/Fe] ratios. The abundance ratios
among Fe-peak elements, e.g., [Cr/Fe] and [Ni/Fe], in metal-poor red giants
exhibit clear trends with little scatter \citep{cayrel04,lai08}, although some
of them simply trace the solar abundance ratio. The small observed scatter in
the [Zn/Fe] ratios is perhaps surprising, given the expected sensitivity of the
yields of Fe-peak nuclei to the parameters used to model explosive
nucleosynthesis. It should also be noted that the heavy neutron-capture elements
(e.g., Sr, Ba, Eu; $Z\geq 38$), which are also thought to be produced during the
explosion of core-collapse supernovae, exhibit a very large scatter in their
abundance ratios \citep[e.g.][]{mcwilliam95b,honda04b}. No clear excess of Zn is
found for objects that show large enhancements of heavy neutron-capture elements
\citep{cayrel04,francois07}, nor of light ones \citep{honda06,honda07}.

Although the [Zn/Fe] ratios in the Galactic halo exhibit a clear trend with
little scatter, a few stars are known that have exceptionally low or high Zn
abundances. \citet{ivans03} investigated three halo stars with unusually low
$\alpha$-element abundances, and reported one with a low [Zn/Fe] abundance
ratio, while the remaining two have quite high values. Moreover, the Zn
abundances of intermediate and higher metallicity stars in dwarf satellites
around the Galaxy appear systematically lower than those of field stars
\citep[e.g.,][]{cohen10}. These observations indicate that further
investigations to understand the origins of Zn, and its implications for the
chemical evolution of the Galaxy (as well as in dwarf galaxies), are strongly
desired.

In this paper we report our measurements of Zn abundances for the sample of
very metal-poor stars studied by \citet{honda04b}. Among them, one star
(BS~16920--017) turned out to have an exceptionally high [Zn/Fe] ratio. The
detailed chemical abundance pattern of this object, determined by an abundance
analysis relative to an object having similar atmospheric parameters (HD~4306),
is also reported. Our observations and measurements are summarized in Section 2.
In Section 3, details of the abundance analyses and the results are reported.
The significance of the Zn excess in BS~16920--017 and its implications are
discussed in Section 4.

\section{Observations and Measurements}\label{sec:obs}

Zinc abundances are determined from the high-resolution spectra reported in
\citet{honda04a,honda04b} (hereafter, Paper I and II, respectively), which were
obtained with the High Dispersion Spectrograph
\citep[HDS; ][]{noguchi02} of the Subaru Telescope
(Table~\ref{tab:obs}). The spectra cover the wavelength range 3500--5100~{\AA},
including the two \ion{Zn}{1} lines at 4722~{\AA} and 4810~{\AA}, with a
resolving power of $R=50,000$. Details of the observations have been reported in
Paper I.

The signal-to-noise ratio of the previous spectrum of {\BS} was not as high as
the average of other stars in the sample of \citet{honda04a}. In addition, the
\ion{Zn}{1} 4810~{\AA} line of this object was affected by a bad column on the
CCD. In order to improve the data quality around the Zn lines, another spectrum
of {\BS} was obtained with the same instrument, covering 4030~{\AA} to
6800~{\AA}, with a resolving power of $R=60,000$, in May 2004. The observation
details are given in Table~\ref{tab:obs}.

Standard data reduction procedures (bias subtraction, flat-fielding,
background subtraction, extraction, and wavelength calibration) are
carried out with the IRAF echelle package\footnote{IRAF is distributed
by the National Optical Astronomy Observatories, which is operated
by the Association of Universities for Research in Astronomy,
Inc. under cooperative agreement with the National Science
Foundation.}, as described in Paper I and \citet[][Paper III]{aoki05}.

Equivalent widths of the two \ion{Zn}{1} lines were measured by the fitting of
Gaussian profiles; results of this exercise are listed in Table \ref{tab:ewzn}.
In order to investigate the relative abundances for {\BS} and {\HD}, equivalent
widths of other elements were measured for a common line set for these two
stars. The line data are adopted from Paper III and \citet[][Paper IV]{aoki07}.
The $gf$ values of Zn I are obtained from Biemont \& Godefroid (1980). In the
lines of this list, the effect of hyperfine splitting in some Mn and Ba lines is
significant. While the effect of hyperfine splitting was included in the Ba
analysis in Paper II, it was not for the Mn lines. Here we take this effect into
account for the Mn analysis, using the line data of \citet{mcwilliam95b}. The
full line list and measured equivalent widths are listed in Table \ref{tab:ew}.

Since the spectral resolution and on-chip binning mode used to obtain the two
spectra of {\BS} are different, they were not merged in the present analysis.
For the abundance analysis from lines in the overlapping wavelength range,
priority is given to previous measurements (Paper I). For {\HD}, we adopted the
equivalent widths measured by \citet{mcwilliam95b} for the red range, which was
not covered by our previous observations. Although the quality of the data of
\citet{mcwilliam95b} is not as high as that of our Subaru spectra in general,
their red spectrum of the bright giant {\HD} has sufficient quality for our
purposes. Our measurements of equivalent widths for {\HD} exhibit no systematic
difference from those of \citet{mcwilliam95b} for the wavelength range in common
(see Fig. 6 of Paper I).

Radial velocities are measured using clean \ion{Fe}{1} lines. The results are
given in Table~\ref{tab:obs}, along with the observing date for each spectrum. No
significant change of heliocentric radial velocity is found for {\BS} between
the two observations in April 2001 and May 2004. Prior to our observations,
\citet{priet00} obtained a radial velocity of this star (--210 $\pm$ 10 km
s$^{-1}$) based on medium-resolution ($R=2000$) spectra. This value
agrees with our result within their reported measurement error. Hence, there
exists no evidence of binarity for {\BS} based on the data obtained thus far.

\section{Abundance Analyses and Results}\label{sec:ana}

As was done in Paper II, chemical abundance analyses are performed using the
analysis tool SPTOOL developed by Y. Takeda (Takeda 2005, private communication),
based on Kurucz's ATLAS9/WIDTH9 program (Kurucz 1993). SPTOOL calculates
synthetic spectra and equivalent widths of lines on the basis of the given
atmospheric parameters, line data, and chemical composition, under the
assumption of LTE.

We adopted the model atmosphere parameters (Table 4; effective temperature,
surface gravity, microturbulent velocity, and metallicity) derived in Paper II, and
listed in Table 2 of that paper. In the previous study, the effective
temperature ({\teff}) was derived from photometric data (primarily $V-K$),
adopting the temperature scale of \citet{alon96}. The microturbulent velocity
({\vt}) and the surface gravity ({\logg}) are determined from the usual
constraint that the abundances derived from individual \ion{Fe}{1} and
\ion{Fe}{2} lines are consistent with one another. The iron abundance ([Fe/H])
is adopted as the metallicity; the [Fe/H] derived from \ion{Fe}{1} and
\ion{Fe}{2} lines are equivalent within the reported errors.

Cayrel et al. (2004) and Roederer et al. (2010) pointed out that the abundance
of Mn derived from the resonance triplet at 4030~{\AA} is systematically lower
($\sim$ 0.4 dex) than the abundance from the other Mn lines. Our results for
the abundance analysis in {\BS} and {\HD} show the same trend. If we also adopt the
correction of 0.4 dex for the abundance from the triplet lines, the result is in
good agreement with the abundance derived from the other Mn lines. In that case,
the derived abundance of Mn in HD~4306 is in agreement with previous studies
(Paper II, McWilliam et al. 1995). However, the cause of this difference remains
as a matter to be discussed further. In this paper, we adopted the mean value of
6 lines with individually large errors.

\subsection{Zinc Abundances for 18 Very Metal-Poor Stars}\label{sec:nazn}

\begin{figure} 
\includegraphics[width=16cm]{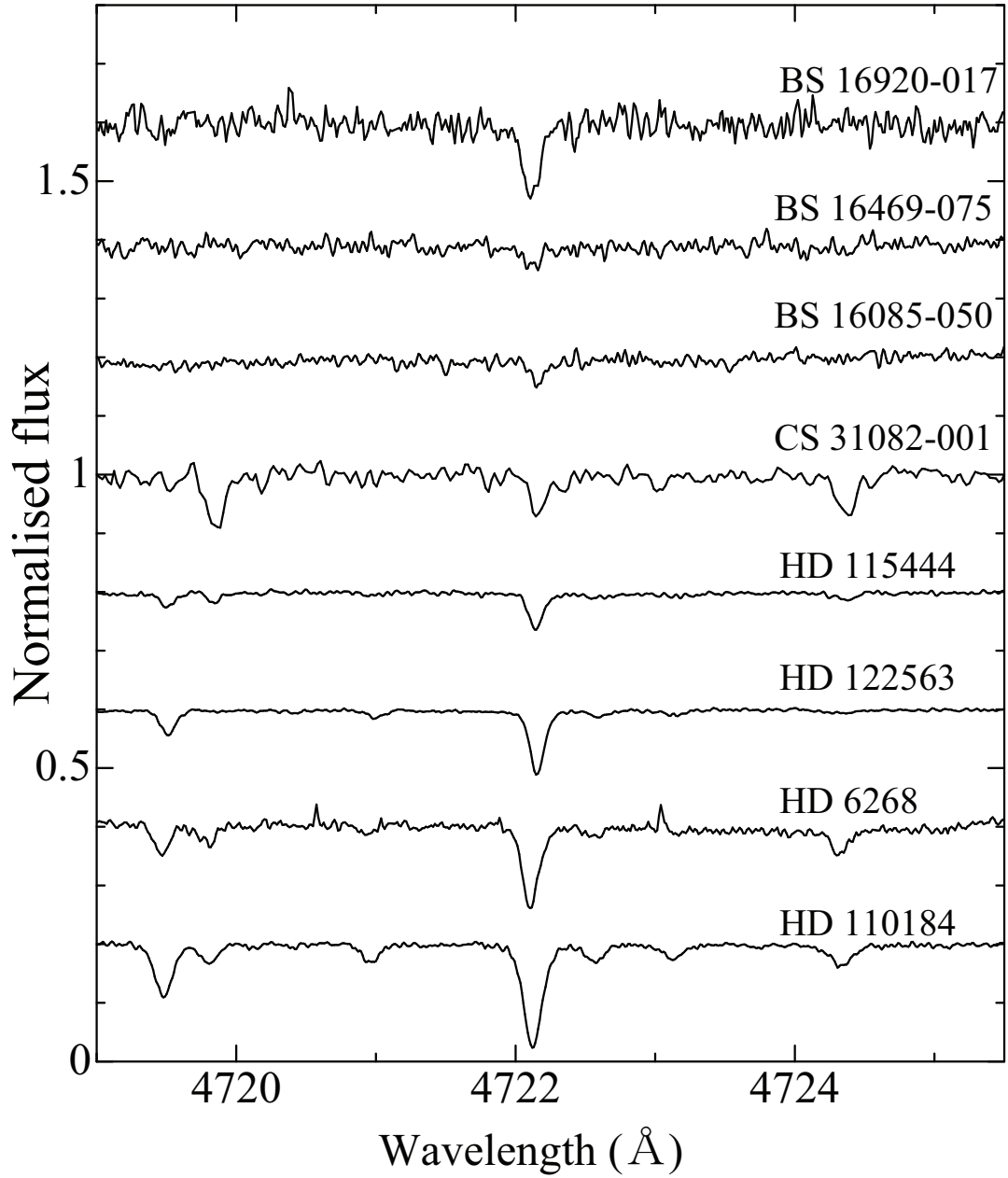} 
\caption[]{Examples of observed spectra in the vicinity of the Zn~I 4722~{\AA}
line. A vertical offset is applied between the spectra.}
\label{fig:spec} 
\end{figure}

We use the two Zn I lines at 4722~{\AA} and 4810~{\AA} for the determination of
Zn abundances; the region of the \ion{Zn}{1} 4722~{\AA} line for several spectra
is shown in Figure \ref{fig:spec}. Contamination from other spectral lines is
not evident at the wavelengths of these two lines in very metal-poor stars. The
measured equivalent widths and the abundances from individual lines are listed
in Table 2. In general, we adopt the straight mean of the results from the two
lines as the final result for the Zn abundance. An exception is CS~30306--132,
for which only the 4722~{\AA} line is covered by our spectrum. The three stars
BS~16082--129, CS~22892--052, and CS~22952--015 exhibit large discrepancies
($\sim$ 0.3 dex) between the abundances derived from the two lines. Moreover,
the equivalent widths of the 4722~{\AA} line are larger than those of the
4810~{\AA} in these three stars, even though the {\loggf} value of the
4810~{\AA} line is greater than that of the other line. 
\citet{sneden03} measured the 4810~{\AA} line of CS~22892--052 to be
8~m{\AA}, which is in agreement with our value (9.2~m{\AA}). 
The equivalent width of the 4810~{\AA} line of CS~22952--015 measured by
\citet{cayrel04} (5.3~m{\AA}) is smaller than our measured value
(8.5~m{\AA}), but this discrepancy can be fully accounted for by the reported
measurement error (3.2~m{\AA} for weak lines, see Paper I).
This gives us confidence in our measurement of the 4810~{\AA} line rather 
than the 4722~{\AA} line. Therefore, for these three stars we adopt the 
Zn abundance from the 4810~{\AA} line as the final result.
Excluding the above three objects, the offset of the mean abundances
from the two spectral lines is only 0.014 $\pm 0.016$~dex, which is much smaller than
the scatter of the abundance differences from the two lines (0.062~dex).

Two sources of errors are included in our estimates of the accuracy for our
derived abundances. One is random errors, which might be caused by the adopted
line data and equivalent width measurements. The size of the random errors are
estimated from the mean of the standard deviations (1 $\sigma$) of the abundances
derived from individual lines for elements that had three or more lines
available. Another source is errors arising from uncertainties in the adopted
atmospheric parameters. The effect on the [Zn/Fe] values is estimated by
changing the atmospheric parameters for {\BS} as follows; $\Delta$ {\teff} =
100~K ($\Delta$[Zn/Fe] = 0.04 dex), $\Delta$ {\vt} = +0.3 dex (0.08 dex), $\Delta$
{\logg} = +0.5 dex (0.02 dex), and $\Delta$ [Fe/H] = +0.3 dex (0.05 dex).

The derived Zn abundances are plotted in Figure 2, along with the results
obtained by previous studies. \citet{sneden91} investigated the Zn abundances of
stars with [Fe/H] $\simeq$ --3, and showed that most objects exhibit solar
ratios of [Zn/Fe]. However, \citet{primas00} investigated Zn abundances for a
larger sample, and found that a few very metal-poor stars exhibit
over-abundances of [Zn/Fe]. This trend becomes clearer from the observations of
\citet{cayrel04} and \citet{nissen07}.  Our study confirms the
increasing trend of [Zn/Fe] with decreasing metallicity for [Fe/H] $<-2.5$. We
note that non-LTE corrections for the Zn lines we have employed are
quantitatively not significant for the stars in our sample \citep{takeda05}.
Energetic core-collapse supernovae (hypernovae) are suggested to be a possible
origin of these Zn excesses (see Section 4).

\begin{figure} 
\includegraphics[angle=-90,width=16cm]{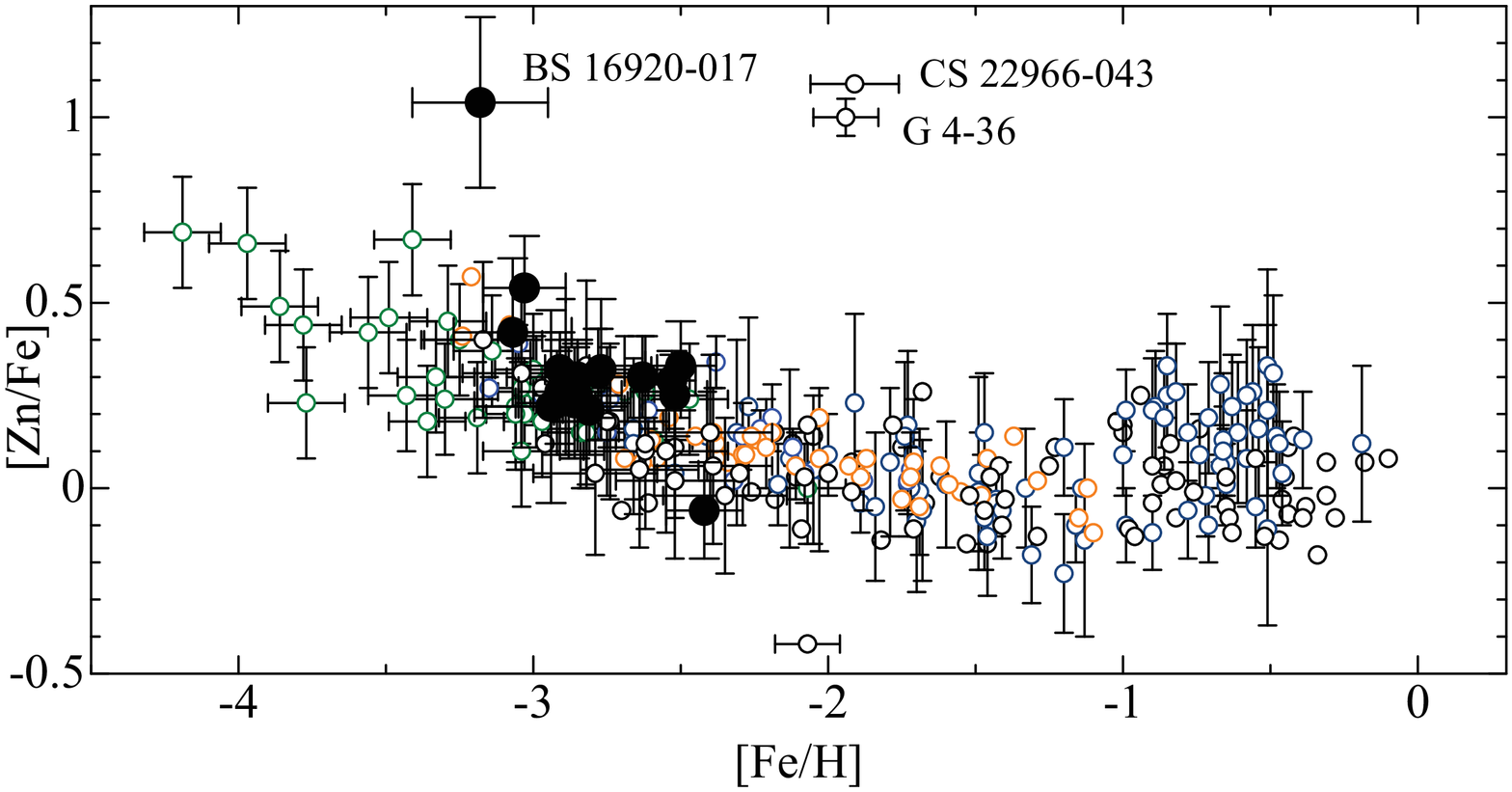} 
\includegraphics[angle=-90,width=16cm]{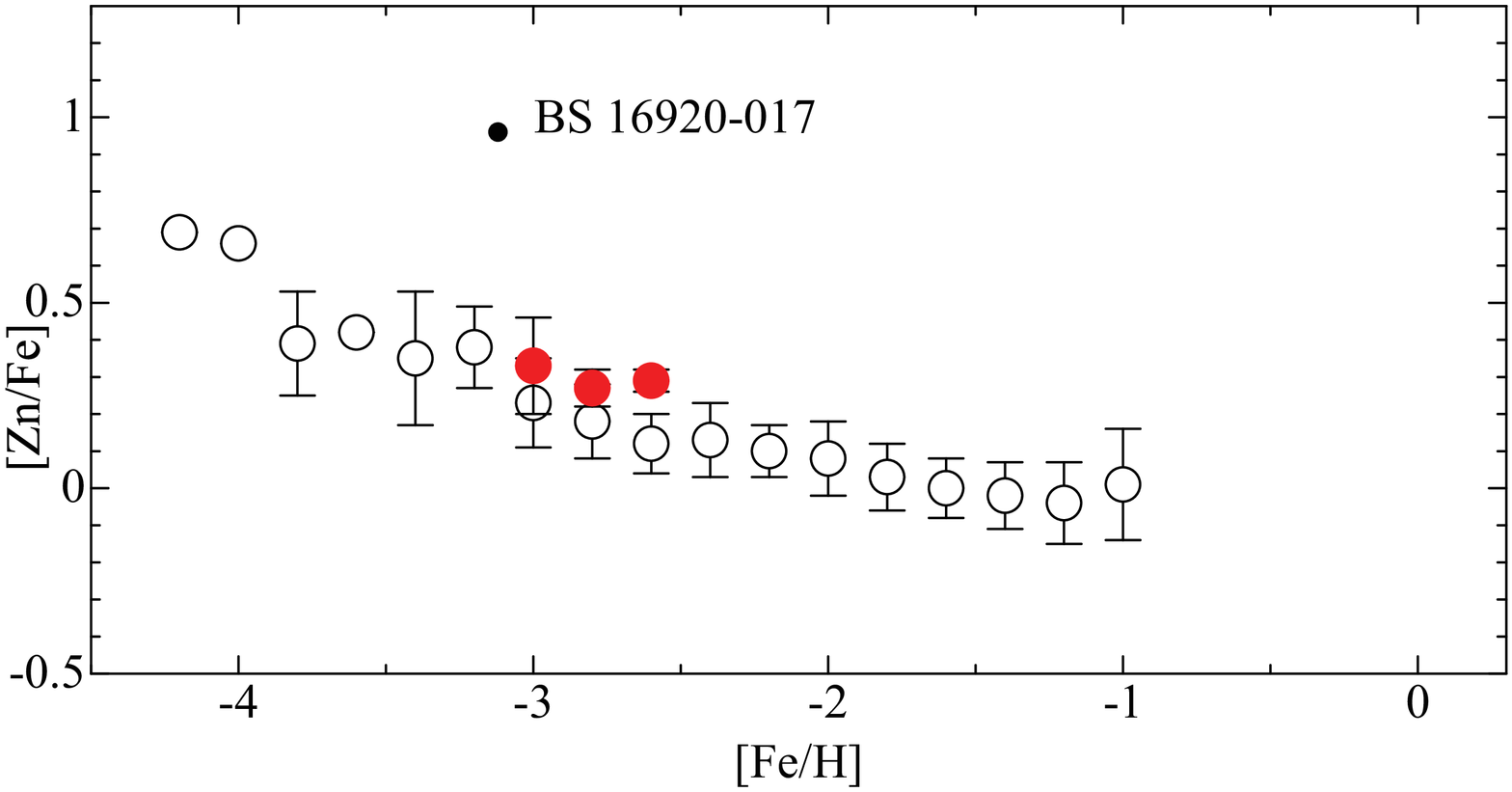} 
\caption[]{{\it Upper panel}: [Zn/Fe] abundance ratios as a function of [Fe/H] 
for our sample stars (filled circles). Results of previous studies are shown by
open circles (Sneden et al. 1991; Johnson 2002; Mishenina et al. 2002; Ivans et
al. 2003; Cayrel et al. 2004; Aoki et al. 2005; Nissen et al. 2007; Saito et al.
2009). {\it Lower panel}: The average [Zn/Fe] abundance ratios in 0.2 dex bins of
[Fe/H] are shown by large filled circles for our sample and by open circles for
those of previous studies. The bar indicates the standard deviation of the [Zn/Fe] values of each bin. The data of {\BS} is not included in the statistics.}
\label{fig:zn} 
\end{figure}

By contrast to the clear trend of [Zn/Fe] with declining [Fe/H] and the small
associated scatter, {\BS} exhibits a very large abundance of Zn ([Zn/Fe] = +1 at
[Fe/H] = --3), higher than reported among very metal-poor stars by previous
studies \citep{cayrel04,nissen07}. In the sample of \citet{cayrel04}, [Zn/Fe]
reaches to $\sim$ +0.7 at [Fe/H]=$-4$. In order to investigate the significance
of the Zn excess in {\BS}, the average and scatter of [Zn/Fe] are calculated for
metallicity bins of $\Delta$[Fe/H] = 0.2 (Figure \ref{fig:zn}). The average
[Zn/Fe] at [Fe/H] = --3.0 is +0.33 with the standard deviation of 0.13 dex in
our sample, while they are +0.23 and 0.12 dex, respectively, for the entire
sample shown in Figure \ref{fig:zn}. These averages are also in agreement with
the result of \citet{saito09} for metallicity bins of $\Delta$ [Fe/H] = 0.5 dex.

At higher metallicity ([Fe/H] $\sim -2$), two halo stars with exceptionally
large Zn abundances were reported by \citet{ivans03} (G~4--36 and CS~22966--043;
[Zn/Fe] $\simeq$+1). Both of these metal-poor stars exhibit low abundances of
their $\alpha$ elements and neutron-capture elements, along with high abundances of
Fe-peak elements. We discuss comparisons of {\BS} with those two stars in
Section 4. 

\subsection{Relative Abundances of {\BS} with respect to {\HD}}\label{sec:rezn}

In an attempt to elucidate the reason for the very high [Zn/Fe] abundance ratio
in {\BS}, we investigate the detailed abundance patterns of other elements for
this object (Table 5). To avoid the systematic errors in the abundance analyses,
such as possible non-LTE effects and uncertainties in the adopted temperature
scale, we select {\HD} as a reference star, and performed abundance analyses
using a common line list. {\HD} is a bright very metal-poor star having similar
stellar parameters to {\BS}, and it exhibits a typical elemental abundance
pattern found in very metal-poor stars (Paper II).

While the difference in metallicity ([Fe/H]) between the two stars is about 0.3
dex, larger differences are seen between other elements. The differences are
also evident from direct inspection of the spectra (see Figures 3-5). The abundance
patterns of the two stars are shown in Figure~\ref{fig:abund}; the lower panel
depicts the abundance differences, on the logarithmic scale, between the two
stars. 

As an overall trend, {\BS} exhibits higher abundances of its iron-peak elements, 
compared with its $\alpha$ elements and neutron-capture elements, relative to {\HD}.  
Mn, Ni, and Zn have remarkably high values compared to Cr, Fe, and Co.  
On the other hand, the abundances of the $\alpha$ elements (Mg, Si, Ca), 
as well as Na and Sc, in {\BS} are deficient compared with those of {\HD}.  
The neutron-capture elements Sr and Ba also exhibit significant under-abundances.  
The Ba abundance of {\BS} is among the lowest values found by previous studies
of very metal-poor stars \citep[e.g.,][]{honda04b,francois07}.

Given the low [$\alpha$/Fe] ratio found in {\BS} with respect to other EMP
stars (see below), it may be interesting to adopt the $\alpha$
elements as the metallicity reference in such comparisons. While most objects
shown in Figure 2 have over-abundances of their $\alpha$ elements ([$\alpha$/Fe]
$\sim +0.3$; Paper II), that of {\BS} is close to the solar ratio
([$\alpha$/Fe]$ = +0.03$), resulting in [$\alpha$/H]$\sim -3$. Excluding {\BS}, the
[Zn/Fe] at [$\alpha$/H]$ = -3.0$ ([Fe/H]$\sim -3.3$) is slightly higher than
that at [Fe/H]$ = -3.0$ on average. However, the large enhancement of Zn in
{\BS} is still evident even at this value of [$\alpha$/H]. We note for
completeness that the Zn excess of {\BS} is more significant in the comparisons
of [Zn/$\alpha$] at [$\alpha$/H]$\sim -3$.

\section{Discussion and Concluding Remarks}\label{sec:disc}

We have derived Zn abundances for 18 very metal-poor stars ([Fe/H] $< -2.5$). We
confirm the increasing trend of [Zn/Fe] with decreasing metallicity, with little
dispersion, as found by previous studies. Although Zn is usually classified as
an iron-peak element, its origin is still not well understood. The
nucleosynthesis of Zn is thought to be the result of complete Si burning and
neutron-capture processing \citep[e.g.,][]{heger02,umeda02}. Recently, the trend
of increasing [Zn/Fe] with decreasing metallicity has been shown to be
consistent with chemical evolution models including the contributions of
hypernovae \citep[e.g.,][]{kobayashi06, tominaga07}. On the other hand,
\citet{heger10} concluded that the elemental abundance ratios of EMP stars are
explained by their supernovae models without a hypernova component, assuming the
production and ejection of Zn in neutrino-driven winds from a proto-neutron star
\citep[e.g.,][]{pruet05} or the accretion disk of a black hole \citep[e.g.,
][]{froehlich06}. However, their models predict lower Co and Zn abundances than
observed, and they will not be enhanced by the ejecta from the innermost layers
\citep{izutani10}.

We found that one star in our sample, {\BS}, exhibits [Zn/Fe] = +1.04,
substantially larger than the average [Zn/Fe] for other stars with metallicity
near [Fe/H] $= -3.0$. The overall elemental abundance pattern of {\BS}, that is,
the low abundance of its $\alpha$ elements and neutron-capture elements, along
with the high abundance of its iron-peak elements, provides a unique constraint
on possible progenitors. We note that a high [Zn/Fe] could be realized by
depletion of Fe onto dust grains, as observed for interstellar matter as well as
damped Lyman $\alpha$ systems. However, no evidence of such chemical
fractionation is seen in the abundance ratio of carbon, which has a lower
condensation temperature than Zn, for BS~16920--017. Measurement of S, which has
a similar condensation temperature to Zn, would be useful for further
confirmation of this point.

The enhancement of iron-peak elements in EMP stars could be explained by
hypernovae models, as argued by several recent studies. However, among the
iron-peak elements, Mn and Ni, as well as Zn, show particularly large excesses.
According to the models of hypernova nucleosynthesis \citep[e.g.,][]{umeda02},
Zn and Co are enhanced, while no excess is expected for Mn (and Cr). Moreover,
it is still unclear if such a very high [Zn/Fe]($\sim +1$) can be explained even
by hypernovae models (Tominaga et al. 2007), although N. Tominaga (private
communication) suggests that a high Zn abundance ratio ([Zn/Fe]$\sim +0.8$)
might be realized with a consistent abundance pattern of other elements of {\BS}
by a very energetic and bright explosion.

The overall abundance pattern of {\BS} is similar to the few halo stars with
exceptionally low abundances of $\alpha$ elements. \citet{ivans03} investigated
the detailed chemical abundances of three stars having [$\alpha$/Fe] $<0$ at
[Fe/H]$\sim -2$. Interestingly, two of them (G~4--36 and CS~22966--043) have
significantly high Zn abundance ratios ([Zn/Fe] $\sim +1$; see Figure 2), while
the other star (BD~+80$^{\circ}$245) possesses an under-abundance of Zn ([Zn/Fe]
$= -0.4$). Such differences of the abundance ratios between these stars
makes it difficult to simply attribute the low abundances of their $\alpha$
elements to large contributions from Type Ia supernovae. The abundance patterns of
{\BS} and the two Zn-enhanced objects studied by \citet{ivans03} are, at least
qualitatively, very similar -- significant under-abundances of elements with
$Z\leq 20$, large enhancements (with respect to Fe) of Mn and Ni as well as Zn,
and deficiencies of heavy neutron-capture elements are among the common
features. An important difference of {\BS} from the two objects in
\citet{ivans03} is its low metallicity ([Fe/H]$=-3.2$). While contributions of
at least several supernovae are expected before the formation of metal-poor
stars with [Fe/H]$\gtrsim -2$, as is the case for G~4--36 and CS~22966--043, yields
from a single supernova are expected to be dominant in the chemical compositions
of such EMP stars in the so-called supernova-induced, low-mass star formation
scenario (Audouse \& Silk 1995; Shigeyama \& Tsujimoto 1998). The peculiar
abundance of BS~16920--017, including the large Zn excess, would also be
explained by a single supernova event.

Under-abundances of $\alpha$ elements are also found in red giants of dwarf
spheroidal galaxies, in particular in the metallicity range [Fe/H] $>-2$. At
lower metallicity, the majority of dwarf galaxy stars seem to have
over-abundances of their $\alpha$ elements \citep[e.g., ][]{cohen10,frebel10,
norris10}. However, the Sextans dwarf galaxy, at least, includes EMP stars with
abundances of the $\alpha$ elements as low as the solar ratio \citep{aoki09}.
The [Zn/Fe] ratio in dwarf galaxy stars is, however, not enhanced, or possibly
deficient, compared to that of the field halo stars \citep[e.g.][]{cohen10}.
This suggests that {\BS} (and probably the Zn-enhanced stars in \citet{ivans03})
had quite different origins from those of the currently surviving dwarf
spheroidal galaxies, even though both show similarly low abundances of their
$\alpha$ elements. We note that Frebel et al. (2010) found a Zn-enhanced EMP
star in the UMa II dwarf galaxy. This star (UMa II-S1 : [Fe/H]$=-3.1$) exhibits
the highest value of [Zn/Fe] (+0.85), after {\BS}, in the abundance studies of
EMP stars. However, according to the authors of that paper, the Zn lines of this
object are somewhat distorted and the abundance could be overestimated. Hence,
this object is excluded in the above comparisons between dwarf spheroidal stars
and {\BS}.

{\BS} exhibits a high Zn abundance, but shows low abundances of neutron-capture
elements. We could not find extreme enhancement of other elements (e.g.,
[Cu/Fe] $<+0.3$, [Eu/Fe] $<0$). Because no neutron-capture element other than Sr
and Ba is detected in {\BS}, we cannot estimate the relative contributions of r-
and s-processes to the neutron-capture elements of this object. The high [Sr/Ba]
ratio (+1.3) possibly indicates a contribution of the so-called ``weak
r-process'' \citep{pfeiffer01,truran02,wanajo06} or the ``Light Element Primary
Process'' \citep[LEPP:][]{travaglio04,montes07}, which has been introduced to
explain the high [Sr/Ba] ratios found in some EMP stars 
\citep[e.g.,][]{mcwilliam95b,burris00}. Such processes producing
 large enhancements of light neutron-capture elements (Sr - Zr) do not, however,
make an excess of Zn. Indeed, very or extremely metal-poor stars with large
enhancements of light neutron-capture elements studied thus far \citep[HD~122563
and HD~88609;][]{honda06, honda07} do not show any excess of Zn. We note that
the recent study of \citet{allen10} suggests that a non-negligible fraction of
the synthesis of Zn is due to the weak s-process, based on observations of Ba
stars in the metallicity range $-0.7 <$ [Fe/H] $<$ +0.12. Yong et al. (2008)
found differences in the Zn and Cu abundances in the mildly metal-poor ([Fe/H]
$\simeq -1.2$) globular clusters M~4 and M~5, in the sense that the abundances
of Zn and Cu in M~4 are enhanced in comparison with those of M~5. Since an
excess of s-process-elements in M~4 is also shown by Yong et al. (2008), this
process may have influenced the synthesis of Zn. However, the s-process is not
expected to contribute significantly to such EMP stars as {\BS}
\citep{kobayashi06}.

In this study, we determined the behavior of Zn in EMP stars, including 
the Zn-enhanced star {\BS}. The abundance pattern of {\BS} is different from the
typical halo EMP stars: it possesses excesses of iron-peak elements, in
particular Mn and Ni as well as Zn, with respect to its $\alpha$ elements and
neutron-capture elements. This peculiar abundance pattern might be produced by
the so-called hypernovae, which are thought to produce large amounts of Zn.
Existence of such Zn-enhanced stars suggests that the Zn abundance can be a
useful indicator in future ``chemical tagging" to distinguish the origins of
halo stars based on wide-field spectroscopic surveys.

\acknowledgments

We would like to thank Dr. N. Tominaga for useful discussions.
We also thank an anonymous referee for many helpful comments.
S.H., W.A., and M.T.-H are supported by a Grant-in-Aid for Science Research from 
MEXT (21740148) and from JSPS (18104003 and 22540255).
T.C.B. acknowledges partial funding of this work from grants
PHY 02-16783 and PHY 08-22648: Physics Frontiers Center/Joint Institute for Nuclear
Astrophysics (JINA), awarded by the U.S. National Science Foundation.

\clearpage
\begin{deluxetable}{@{}l@{\extracolsep{\fill}}c@{\extracolsep{\fill}}c@{\extracolsep{\fill}}c@{\extracolsep{\fill}}c@{\extracolsep{\fill}}c@{\extracolsep{\fill}}c@{\extracolsep{\fill}}c}
\tablewidth{0pt}
\tablecaption{\label{tab:obs} PROGRAM STARS AND OBSERVATIONS}
\tablehead{
Star    & Wavelength &  Exp.\tablenotemark{a} & S/N\tablenotemark{b}& &  Obs. date (JD) & Radial velocity ~ & \\
        & ({\AA})   &     (minutes)           & (4000~{\AA})&(5000~{\AA})      &                 & (km s$^{-1}$)     &    
}
\startdata
{\BS}~~ & 3500--5100 & 90 (3) &  41/1 ~ & 59/1 ~& 27 Jan.2001  (2451937)  & $-206.5 \pm 0.8$   &   \\
{\BS}~~ & 4030--6800 & 30 (1) &  40/1\tablenotemark{c}~ &62/1\tablenotemark{c}~& 31 May 2004  (2453156)  & $  -206.2 \pm 0.4  $   &   \\
{\HD}   & 3500--5100 & 30 (2) & 272/1 ~ &340/1 ~& 19 Aug.2000  (2451776)  & $ -69.7 \pm 0.3$   &   \\
\enddata
\tablenotetext{a}{Total exposure time (number of exposures).}
\tablenotetext{b}{$S/N$ ratio per pixel estimated from the photon counts at around 4000 and 5000~{\AA}. }
\tablenotetext{c}{2$\times$2 binning mode was used.}
\end{deluxetable}

\begin{deluxetable}{lccccccc}
\tablewidth{0pt}
\tablecaption{EQUIVALENT WIDTHS OF ZN LINES AND DERIVED ABUNDANCES\label{tab:ewzn}} 
\tablehead{
     & \ion{Zn}{1} $\lambda 4722$ & {\loge} &\ion{Zn}{1} $\lambda 4810$ & {\loge} & {\loge}(avg.) & [Zn/Fe] & $\sigma$ \\
}
\startdata
     wavelength({\AA}) & 4722.16 & &4810.54 & &&\\
     {\loggf} & --0.390 && --0.170 & \\
     L.E.P.(eV) &4.030  && 4.080 &\\\hline
     HD~4306       & 7.8  & 2.00 & 11.6 & 2.02 & 2.01 & +0.30 & 0.09 \\
     HD~6268       & 21.3 & 2.29 & 23.4 & 2.17 & 2.23 & +0.30 & 0.11 \\
     HD~88609      & 10.5 & 1.92 & 14.2 & 1.90 & 1.91 & +0.42 & 0.20 \\
     HD~110184     & 29.8 & 2.31 & 35.6 & 2.27 & 2.29 & +0.25 & 0.11 \\
     HD~115444     & 8.8  & 1.97 & 11.7 & 1.94 & 1.96 & +0.25 & 0.15 \\
     HD~122563     & 15.1 & 2.11 & 20.5 & 2.10 & 2.11 & +0.32 & 0.19 \\
     HD~126587     & 7.3  & 2.07 & 11.0 & 2.10 & 2.09 & +0.31 & 0.12 \\
     HD~140283     & 3.1  & 2.30 & 5.0  & 2.34 & 2.32 & +0.29 & 0.08 \\
     HD~186478     & 20.6 & 2.41 & 25.6 & 2.36 & 2.39 & +0.33 & 0.12 \\
     BS~16082--129  & 11.6* & 2.23* & 8.9  & 1.93 & 1.93 & +0.23 & 0.15 \\
     BS~16085--050  & 7.6  & 2.04 & 8.0  & 1.89 & 1.97 & +0.32 & 0.10 \\
     BS~16469--075  & 7.3  & 2.03 & 11.9 & 2.10 & 2.07 & +0.54 & 0.14 \\
     BS~16920--017  & 20.5 & 2.41 & 27.5 & 2.43 & 2.42 & +1.04 & 0.23 \\
     BS~16928--053  & 10.1 & 1.90 & 14.5 & 1.92 & 1.91 & +0.26 & 0.14 \\
     CS~22892--052  & 14.1* & 2.25* & 9.2  & 1.87 & 1.87 & +0.23 & 0.14 \\
     CS~22952--015  & 11.2* & 2.16* & 8.5  & 1.84 & 1.84 & +0.22 & 0.26 \\
     CS~30306--132  & 13.6 & 2.08 & --   & --   & 2.08 & --0.06 & 0.13 \\
     CS~31082--001  & 8.3  & 1.94 & 12.2 & 1.97 & 1.96 & +0.21 & 0.12 \\

\enddata
\tablenotetext{*}{These values are not used to derive the final results.}
\end{deluxetable}

\begin{deluxetable}{lcccccccc}
\tablewidth{0pt}
\tablecaption{EQUIVALENT WIDTHS\label{tab:ew}} 
\tablehead{
Species    & wavelength & {\loggf} & L.E.P. & {\HD} & {\loge} &{\BS}&{\loge}& Remarks 
}
\startdata
     Na I  &  5889.95 &   0.101 &  0.000  &    152 \tablenotemark{a} & 3.94 &  109.6 \tablenotemark{b} &  3.27 \\
     Na I  &  5895.92 &  -0.197 &  0.000  &    138 \tablenotemark{a} & 4.00 &   93.6 \tablenotemark{b} &  3.21 \\
     Mg I  &  3829.35 &  -0.210 &  2.709  &   157.6    & 5.28 &  110.3     &  4.72 \\
     Mg I  &  3832.31 &   0.140 &  2.712  &   190.1    & 5.25 &  103.6     &  4.21 \\
     Mg I  &  3838.29 &   0.414 &  2.717  &   237.7    & 5.25 &  137.1     &  4.65 \\
     Mg I  &  4571.10 &  -5.588 &  0.000  &    45.5    & 5.25 &   17.5     &  4.64 \\
     Mg I  &  5172.68 &  -0.380 &  2.712  &   166 \tablenotemark{a}  & 5.00 &  131.6     &  4.80 \\
     Mg I  &  5183.60 &  -0.160 &  2.717  &   189 \tablenotemark{a}  & 4.98 &  149.7     &  4.86 \\
     Mg I  &  5528.41 &  -0.490 &  4.346  &    52 \tablenotemark{a}  & 5.18 &   28.8 \tablenotemark{b} &  4.75 \\
     Al I  &  3961.53 &  -0.340 &  0.010  &   106.9    & 3.09 &   85.1     &  2.66 \\
     Si I  &  4102.94 &  -2.910 &  1.910  &    63.0    & 5.08 &   22.9     &  4.23 \\
     Ca I  &  4283.01 &  -0.220 &  1.886  &    63 \tablenotemark{a}  & 4.29 &   12.8     &  3.09 \\
     Ca I  &  4318.65 &  -0.210 &  1.899  &    47.4    & 3.96 &   25.2     &  3.48 \\
     Ca I  &  4425.44 &  -0.360 &  1.879  &    42.7    & 3.98 &   11.2     &  3.14 \\
     Ca I  &  4454.78 &   0.260 &  1.899  &    70.3    & 3.96 &   30.8     &  3.13 \\
     Ca I  &  5588.75 &   0.358 &  2.526  &    31 \tablenotemark{a}  & 3.69 &   13.2 \tablenotemark{b} &  3.18 \\
     Ca I  &  5594.47 &   0.097 &  2.523  &    --      &  --  &   10.5 \tablenotemark{b} &  3.32 \\
     Ca I  &  5598.49 &  -0.087 &  2.521  &    --      &  --  &    7.2 \tablenotemark{b} &  3.32 \\
     Ca I  &  6102.72 &  -0.770 &  1.879  &    24 \tablenotemark{a}  & 3.88 &   11.6 \tablenotemark{b} &  3.47 \\
     Ca I  &  6162.17 &   0.100 &  1.899  &    74 \tablenotemark{a}  & 4.01 &   31.1 \tablenotemark{b} &  3.19 \\
     Ca I  &  6439.08 &   0.390 &  2.526  &    40 \tablenotemark{a}  & 3.79 &   17.1 \tablenotemark{b} &  3.25 \\
     Sc II &  4400.40 &  -0.540 &  0.600  &    56.2    & 0.44 &   27.5 \tablenotemark{b} & -0.34 \\
     Sc II &  4415.56 &  -0.670 &  0.595  &    48.9    & 0.42 &   29.2     & -0.18 \\
     Sc II &  5031.02 &  -0.400 &  1.357  &    20.3    & 0.39 &    8.4     & -0.27 \\
     Ti I  &  3998.64 &   0.000 &  0.048  &    54.8    & 2.36 &    --      &  --   \\
     Ti I  &  4533.24 &   0.532 &  0.848  &    38.6    & 2.35 &   30.0     &  2.16 \\
     Ti I  &  4534.78 &   0.336 &  0.836  &     26 \tablenotemark{a} & 2.26 &   19.3     &  2.06 \\
     Ti I  &  4535.57 &   0.120 &  0.826  &     17 \tablenotemark{a} & 2.22 &   14.7     &  2.12 \\
     Ti I  &  4981.73 &   0.560 &  0.848  &    44.5    & 2.39 &   39.2     &  2.29 \\
     Ti I  &  4991.07 &   0.436 &  0.836  &    40.7    & 2.43 &   17.8     &  1.88 \\
     Ti I  &  4999.50 &   0.306 &  0.826  &    35.8    & 2.45 &   25.3     &  2.20 \\
     Ti I  &  5039.96 &  -1.130 &  0.020  &    17.3    & 2.48 &   12.8     &  2.30 \\
     Ti I  &  5064.65 &  -0.935 &  0.048  &    20.5    & 2.41 &   18.9 \tablenotemark{b} &  2.34 \\
     Ti I  &  5173.74 &  -1.062 &  0.000  &    --      & --   &   14.8     &  2.27 \\
     Ti I  &  5192.97 &  -0.948 &  0.021  &    17 \tablenotemark{a}  & 2.28 &   15.5     &  2.20 \\
     Ti I  &  5210.38 &  -0.828 &  0.048  &    --      & --   &   17.8 \tablenotemark{b} &  2.19 \\
     Ti II &  4028.36 &  -1.000 &  1.892  &    38.4    & 2.52 &   26.8     &  2.07 \\
     Ti II &  4337.88 &  -1.130 &  1.080  &    84.5    & 2.71 &   70.2     &  2.26 \\
     Ti II &  4394.07 &  -1.590 &  1.220  &    42.3    & 2.35 &   28.6 \tablenotemark{b} &  1.87 \\
     Ti II &  4395.85 &  -1.970 &  1.243  &    30.6    & 2.52 &   25.1 \tablenotemark{b} &  2.19 \\
     Ti II &  4399.79 &  -1.270 &  1.237  &    74.5    & 2.75 &   57.3 \tablenotemark{b} &  2.22 \\
     Ti II &  4417.72 &  -1.430 &  1.165  &    73.9    & 2.81 &   63.4     &  2.44 \\
     Ti II &  4443.78 &  -0.700 &  1.080  &    96.8    & 2.58 &   87.7     &  2.31 \\
     Ti II &  4450.50 &  -1.510 &  1.084  &    64.7    & 2.56 &   58.1     &  2.28 \\
     Ti II &  4464.46 &  -2.080 &  1.161  &    44.8    & 2.81 &   38.5     &  2.50 \\
     Ti II &  4468.52 &  -0.600 &  1.131  &    98.4    & 2.57 &   89.3     &  2.31 \\
     Ti II &  4470.84 &  -2.280 &  1.165  &    28.1    & 2.67 &   24.5     &  2.38 \\
     Ti II &  4501.27 &  -0.760 &  1.116  &    93.9    & 2.58 &   83.0     &  2.26 \\
     Ti II &  4571.96 &  -0.530 &  1.572  &    85.8    & 2.66 &   76.8     &  2.37 \\
     Ti II &  4589.92 &  -1.790 &  1.237  &    48.4    & 2.66 &   41.6     &  2.35 \\
     Ti II &  4865.61 &  -2.810 &  1.116  &    11.3    & 2.61 &   10.3 \tablenotemark{b} &  2.35 \\
     Ti II &  5129.16 &  -1.390 &  1.892  &    31 \tablenotemark{a}  & 2.64 &   20.6     &  2.19 \\
     Ti II &  5185.90 &  -1.350 &  1.893  &    22 \tablenotemark{a}  & 2.39 &   17.9     &  2.07 \\
     Ti II &  5188.69 &  -1.210 &  1.582  &    54 \tablenotemark{a}  & 2.53 &   49.0     &  2.27 \\
     Ti II &  5226.54 &  -1.300 &  1.566  &    44 \tablenotemark{a}  & 2.41 &   45.9 \tablenotemark{b} &  2.27 \\
     Ti II &  5336.78 &  -1.630 &  1.582  &    30 \tablenotemark{a}  & 2.47 &   27.3 \tablenotemark{b} &  2.21 \\
     V   I &  4379.23 &   0.550 &  0.301  &    14.2    & 1.08 &   5.6 \tablenotemark{b}  &  0.59 \\
     V  II &  3951.96 &  -0.784 &  1.476  &    17.7    & 1.22 &   --       &  -- \\
     V  II &  4005.71 &  -0.522 &  1.820  &    18.1    & 1.37 &   17.6     &  1.15 \\
     Cr  I &  4254.33 &  -0.114 &  0.000  &     82 \tablenotemark{a} & 2.28 &   82.9     &  2.50 \\
     Cr  I &  4274.81 &  -0.321 &  0.000  &   103 \tablenotemark{a}  & 3.08 &   79.6     &  2.59 \\
     Cr  I &  4289.72 &  -0.360 &  0.000  &    85.9    & 2.63 &   72.9     &  2.41 \\
     Cr  I &  5204.51 &  -0.208 &  0.941  &    --      & --   &   67.0 \tablenotemark{b} &  3.00 \\
     Cr  I &  5206.04 &   0.019 &  0.941  &    63 \tablenotemark{a}  & 2.60 &   58.9 \tablenotemark{b} &  2.56 \\
     Cr  I &  5208.44 &   0.160 &  0.940  &    76 \tablenotemark{a}  & 2.75 &   68.8 \tablenotemark{b} &  2.68 \\
     Cr II &  4558.65 &  -0.660 &  4.070  &   10.2     & 3.00 &    8.9     &  2.72 \\
     Cr II &  4588.20 &  -0.630 &  4.070  &   8.4      & 2.87 &    4.1     &  2.32 \\
     Mn  I &  4030.75 &  -0.470 &  0.000  &   91.0     & 1.50 &   98.7     &  1.69 \\
     Mn  I &  4033.06 &  -0.618 &  0.000  &   77.4     & 1.51 &   86.7     &  1.73 \\
     Mn  I &  4034.48 &  -0.811 &  0.000  &   79.1     & 1.83 &   80.8     &  1.92 \\
     Mn  I &  4041.36 &   0.285 &  2.114  &   16.6     & 2.05 &   31.8     &  2.45  & no HF\\
     Mn  I &  4754.05 &  -0.086 &  2.282  &    6.3     & 2.07 &   13.5     &  2.39 \\
     Mn  I &  4823.53 &   0.144 &  2.319  &    8.8     & 2.04 &   13.3     &  2.21 \\
     Fe  I &  3763.80 &  -0.240 &  0.990  & 139.0      & 4.59 & 98.0       & 4.03 \\ 
     Fe  I &  3767.19 &  -0.390 &  1.010  & 125.9      & 4.58 & 90.8       & 3.99 \\
     Fe  I &  3787.88 &  -0.859 &  1.011  & 112.0      & 4.76 & 85.3       & 4.25 \\
     Fe  I &  3805.34 &   0.310 &  3.301  & 46.8       & 4.41 &  --        & -- \\
     Fe  I &  3815.84 &   0.226 &  1.485  & 137.2      & 4.62 & 96.9       & 4.06 \\
     Fe  I &  3820.43 &   0.120 &  0.860  & 188.2      & 4.46 & 149.9      & 4.30 \\
     Fe  I &  3825.88 &  -0.024 &  0.920  & 165.1      & 4.51 & 131.0      & 4.28 \\
     Fe  I &  3827.82 &   0.062 &  1.557  & 124.3      & 4.68 & 107.9      & 4.58 \\
     Fe  I &  3840.44 &  -0.506 &  0.990  & 123.9      & 4.56 & 92.1       & 4.04 \\
     Fe  I &  3849.98 &  -0.863 &  1.010  & 110.6      & 4.67 & 99.5       & 4.62 \\
     Fe  I &  3856.37 &  -1.280 &  0.052  & 137.6      & 4.67 & 123.1      & 4.71 \\
     Fe  I &  3859.91 &  -0.710 &  0.000  & 184.0      & 4.53 & 142.5      & 4.40 \\
     Fe  I &  3865.52 &  -0.982 &  1.011  & 111.2      & 4.79 & 90.4       & 4.47 \\
     Fe  I &  3886.28 &  -1.080 &  0.050  & 115.4      & 4.00 & 125.8      & 4.56 \\
     Fe  I &  3899.71 &  -1.531 &  0.087  & 127.7      & 4.75 & 118.0      & 4.88 \\
     Fe  I &  3902.95 &  -0.466 &  1.557  & 103.8      & 4.73 & 92.4       & 4.66 \\
     Fe  I &  3920.26 &  -1.746 &  0.121  & 121.8      & 4.87 & 114.1      & 5.02 \\
     Fe  I &  3922.91 &  -1.651 &  0.052  & 128.9      & 4.84 & 114.2      & 4.85 \\
     Fe  I &  3949.95 &  -1.250 &  2.176  & 45.5       & 4.61 & 26.8       & 4.18 \\
     Fe  I &  4005.24 &  -0.610 &  1.557  & 101.5      & 4.79 & 85.0       & 4.54 \\
     Fe  I &  4063.59 &   0.060 &  1.557  & 128.4      & 4.68 & 110.3      & 4.57 \\
     Fe  I &  4071.74 &  -0.022 &  1.608  & 120.6      & 4.67 & 98.0 \tablenotemark{b}   & 4.38 \\
     Fe  I &  4076.63 &  -0.530 &  3.210  & 26.7       & 4.64 & 12.4 \tablenotemark{b}   & 4.18  \\
     Fe  I &  4114.45 &  -1.300 &  2.832  & 12.8       & 4.55 &  --        & -- \\
     Fe  I &  4132.90 &  -1.010 &  2.845  & 23.1       & 4.61 & 12.4 \tablenotemark{b}   & 4.24  \\
     Fe  I &  4143.87 &  -0.510 &  1.557  & 104.0      & 4.69 & 102.3      & 4.91 \\
     Fe  I &  4147.67 &  -2.104 &  1.485  & 47.5       & 4.66 & 30.9       & 4.49 \\
     Fe  I &  4154.50 &  -0.690 &  2.832  & 38.1       & 4.62 & 15.7 \tablenotemark{b}   & 4.02 \\
     Fe  I &  4156.81 &  -0.810 &  2.832  & 39.6       & 4.77 & 17.5 \tablenotemark{b}   & 4.20 \\
     Fe  I &  4157.78 &  -0.400 &  3.417  & 27.8       & 4.77 &  --        & -- \\
     Fe  I &  4174.91 &  -2.970 &  0.920  & 41.9       & 4.73 & 30.9       & 4.49 \\
     Fe  I &  4175.64 &  -0.830 &  2.845  & 30.1       & 4.60 &  --        & -- \\
     Fe  I &  4181.76 &  -0.370 &  2.832  & 53.2       & 4.63 & 38.7 \tablenotemark{b}   & 4.32 \\
     Fe  I &  4182.38 &  -1.180 &  3.017  & 8.6        & 4.44 &   --  & -- \\
     Fe  I &  4187.04 &  -0.548 &  2.450  & 64.7       & 4.64 & 44.4       & 4.20 \\
     Fe  I &  4187.79 &  -0.550 &  2.420  & 66.5       & 4.66 & 54.6       & 4.43 \\
     Fe  I &  4191.43 &  -0.670 &  2.469  & 54.7       & 4.54 & 41.8       & 4.28 \\
     Fe  I &  4195.33 &  -0.490 &  3.332  & 30.5       & 4.82 & 7.8 \tablenotemark{b}    & 4.04 \\
     Fe  I &  4199.10 &   0.160 &  3.047  & 61.0       & 4.54 & 38.0       & 4.03 \\
     Fe  I &  4202.03 &  -0.689 &  1.490  & 102.3      & 4.74 & 94.2       & 4.76 \\
     Fe  I &  4222.21 &  -0.967 &  2.450  & 45.6       & 4.60 & 29.9       & 4.26 \\
     Fe  I &  4227.43 &   0.270 &  3.332  & --         & --   & 40.2       & 4.30 \\
     Fe  I &  4233.60 &  -0.604 &  2.482  & --         & --   & 58.9       & 4.67 \\
     Fe  I &  4250.12 &  -0.405 &  2.469  & --         & --   & 46.7       & 4.12 \\
     Fe  I &  4250.79 &  -0.710 &  1.557  & --         & --   & 87.6       & 4.63 \\
     Fe  I &  4260.47 &   0.080 &  2.399  & --         & --   & 79.9       & 4.55 \\
     Fe  I &  4271.15 &  -0.349 &  2.450  & --         & --   & 52.3 \tablenotemark{b}   & 4.17 \\
     Fe  I &  4325.76 &   0.010 &  1.608  & 125.6      & 4.67 & 123.8      & 4.89 \\
     Fe  I &  4337.05 &  -1.695 &  1.557  & 62.1       & 4.64 & 40.4 \tablenotemark{b}   & 4.16 \\
     Fe  I &  4383.56 &   0.208 &  1.490  & 143.6      & 4.58 & 127.8 \tablenotemark{b}  & 4.58 \\
     Fe  I &  4404.76 &  -0.147 &  1.560  & 123.3      & 4.68 & 101.5 \tablenotemark{b}  & 4.41 \\
     Fe  I &  4415.14 &  -0.621 &  1.610  & 104.0      & 4.77 & 100.3      & 4.90 \\
     Fe  I &  4430.61 &  -1.659 &  2.223  & 26.7       & 4.60 &  --        & -- \\
     Fe  I &  4442.34 &  -1.255 &  2.198  & 49.3       & 4.64 & 32.8       & 4.29 \\
     Fe  I &  4443.19 &  -1.040 &  2.858  & 18.9       & 4.51 & 16.5 \tablenotemark{b}   & 4.41 \\
     Fe  I &  4447.72 &  -1.342 &  2.223  & 44.7       & 4.66 & 23.0       & 4.16 \\
     Fe  I &  4466.55 &  -0.600 &  2.832  & 49.3       & 4.74 & 28.6       & 4.28 \\
     Fe  I &  4489.74 &  -3.966 &  0.121  & 40.3       & 4.70 & 28.0 \tablenotemark{b}   & 4.41 \\
     Fe  I &  4494.56 &  -1.136 &  2.198  & 54.9       & 4.64 & 41.6       & 4.37 \\
     Fe  I &  4528.61 &  -0.822 &  2.176  & 71.0       & 4.66 & 57.2       & 4.40 \\
     Fe  I &  4531.15 &  -2.155 &  1.485  & 48.9       & 4.69 & 31.7       & 4.32 \\
     Fe  I &  4602.94 &  -2.210 &  1.485  & 46.3       & 4.68 & 30.8       & 4.34 \\
     Fe  I &  4736.77 &  -0.750 &  3.210  & 23.1       & 4.72 & 12.2       & 4.35 \\
     Fe  I &  4871.32 &  -0.362 &  2.870  & 51.2       & 4.53 & 37.6       & 4.26 \\
     Fe  I &  4872.14 &  -0.570 &  2.882  & 41.5       & 4.56 & 29.7       & 4.30 \\
     Fe  I &  4890.76 &  -0.390 &  2.876  & 51.0       & 4.56 & 32.3       & 4.17 \\
     Fe  I &  4891.49 &  -0.110 &  2.851  & 64.5       & 4.55 & 48.8       & 4.24 \\
     Fe  I &  4903.31 &  -0.930 &  2.882  & 25.6       & 4.57 & 15.3       & 4.26 \\
     Fe  I &  4918.99 &  -0.340 &  2.865  & 52.7       & 4.53 & 37.1       & 4.22 \\
     Fe  I &  4924.77 &  -2.256 &  2.279  & 10.1       & 4.69 & 6.8 \tablenotemark{b}    & 4.46 \\
     Fe  I &  4938.81 &  -1.080 &  2.876  & 18.5       & 4.52 & 12.3       & 4.28 \\
     Fe  I &  4939.69 &  -3.340 &  0.859  & 30.0       & 4.69 & 20.8       & 4.44 \\
     Fe  I &  4946.39 &  -1.170 &  3.368  & 7.2        & 4.70 &  --   & -- \\
     Fe  I &  4966.09 &  -0.870 &  3.332  & 13.9       & 4.68 & 8.7        & 4.42 \\
     Fe  I &  4973.10 &  -0.950 &  3.960  & 3.8        & 4.86 &  --   & -- \\
     Fe  I &  4994.13 &  -2.956 &  0.915  & 39.8       & 4.57 & 31.0       & 4.37 \\
     Fe  I &  5001.87 &   0.050 &  3.880  & 16.6       & 4.49 & 10.6       & 4.23 \\
     Fe  I &  5006.12 &  -0.610 &  2.833  & 42.5       & 4.54 & 28.4       & 4.24 \\
     Fe  I &  5014.94 &  -0.300 &  3.943  & 10.2       & 4.65 &  --        & -- \\
     Fe  I &  5022.24 &  -0.530 &  3.984  & 6.4        & 4.71 &  --   & -- \\
     Fe  I &  5041.76 &  -2.200 &  1.485  & 52.1       & 4.73 & 33.2       & 4.34 \\
     Fe  I &  5044.21 &  -2.040 &  2.850  &  3.8       & 4.66 &  --   & -- \\
     Fe  I &  5049.82 &  -1.344 &  2.279  & 39.5       & 4.56 & 27.0  & 4.28 \\
     Fe  I &  5051.64 &  -2.795 &  0.915  & 56.9       & 4.74 & 44.0  & 4.49 \\
     Fe  I &  5068.77 &  -1.040 &  2.940  & 17.1       & 4.50 & 14.2  & 4.38 \\
     Fe  I &  5074.75 &  -0.200 &  4.220  & 8.0        & 4.75 &  --   & -- \\
     Fe  I &  5083.34 &  -2.958 &  0.958  & 44.1       & 4.70 & 31.3  & 4.42 \\
     Fe  I &  5090.77 &  -0.360 &  4.256  & 4.5        & 4.67 &  --   & -- \\
     Fe  I &  5123.73 &  -3.068 &  1.011  & --     & --   & 26.6  & 4.48 \\
     Fe  I &  5127.37 &  -3.307 &  0.915  & --     & --   & 17.6  & 4.36 \\
     Fe  I &  5150.84 &  -3.003 &  0.990  & --     & --   & 20.2  & 4.22 \\
     Fe  I &  5151.91 &  -3.322 &  1.011  & --     & --   & 16.8  & 4.47 \\
     Fe  I &  5162.27 &   0.020 &  4.178  & --     & --   & 10.1  & 4.57 \\
     Fe  I &  5171.60 &  -1.793 &  1.485  & --     & --   & 58.3  & 4.49 \\
     Fe  I &  5191.46 &  -0.550 &  3.038  & --     & --   & 19.7  & 4.18 \\
     Fe  I &  5192.34 &  -0.420 &  2.998  & --     & --   & 26.0  & 4.17 \\
     Fe  I &  5194.94 &  -2.090 &  1.557  & --     & --   & 37.8  & 4.40 \\
     Fe II &  4178.86 &  -2.480 &  2.583  & 41.2   & 4.77 & 21.3 \tablenotemark{b}  &  4.11\\
     Fe II &  4233.17 &  -2.000 &  2.583  & --     & --   & 42.9 \tablenotemark{b}  & 4.15 \\
     Fe II &  4416.82 &  -2.600 &  2.778  & 21.2   & 4.63 & 10.6 \tablenotemark{b}  & 4.05 \\
     Fe II &  4491.40 &  -2.700 &  2.856  & 15.8   & 4.66 & 8.6   & 4.13 \\
     Fe II &  4508.28 &  -2.580 &  2.856  & 26.6   & 4.84 & 18.4  & 4.41 \\
     Fe II &  4515.34 &  -2.480 &  2.844  & 22.3   & 4.61 & 16.0  & 4.22 \\
     Fe II &  4520.22 &  -2.600 &  2.810  & 21.3   & 4.67 & 12.3  & 4.16 \\
     Fe II &  4541.52 &  -3.050 &  2.856  & 7.7    & 4.64 &  --   & -- \\
     Fe II &  4555.89 &  -2.290 &  2.828  & 26.7   & 4.51 &  --   & -- \\
     Fe II &  4576.33 &  -2.940 &  2.844  & 9.0    & 4.59 &  --   & -- \\
     Fe II &  4583.83 &  -2.020 &  2.807  & 53.2   & 4.78 & 41.9  & 4.38 \\
     Fe II &  4923.93 &  -1.320 &  2.891  & 73.7   & 4.58 & 70.0  & 4.41 \\
     Fe II &  4993.35 &  -3.670 &  2.810  & 1.8    & 4.50 &  --   & -- \\
     Fe II &  5018.44 &  -1.292 &  2.890  & 84.4   & 4.78 & 77.3  & 4.55 \\
     Co  I &  3842.05 &  -0.770 &  0.923  &    32.9 & 2.24 &   27.9 & 2.10 \\
     Co  I &  3845.47 &   0.010 &  0.923  &    63.3 & 2.19 &   48.2 & 1.84 \\
     Co  I &  3873.12 &  -0.660 &  0.432  &    96.0 & 3.33 & 63.2 & 2.40 & not used\\
     Co  I &  3881.87 &  -1.130 &  0.582  &    43.2 & 2.42 &   38.7 & 2.32 \\
     Co  I &  3995.31 &  -0.220 &  0.923  &    59.0 & 2.26 &   49.3 & 2.07 \\
     Co  I &  4020.90 &  -2.070 &  0.432  &    13.7 & 2.40 &   --   & -- \\
     Co  I &  4110.53 &  -1.080 &  1.049  &    12.6 & 2.09 &   --   & -- \\
     Co  I &  4121.32 &  -0.320 &  0.923  &    62.9 & 2.43 &   54.9 & 2.30 \\
     Ni  I &  3807.14 &  -1.220 &  0.423  &    85.3 & 3.39 &   83.1 & 3.56 \\
     Ni  I &  3858.30 &  -0.951 &  0.423  &    93.0 & 3.34 &   89.5 & 3.48 \\
     Ni  I &  5476.90 &  -0.890 &  1.826  &     45 \tablenotemark{a} & 3.32 &   51.4 \tablenotemark{b} & 3.46 \\
     Ni  I &  6643.64 &  -2.300 &  1.676  &     --  & --     &    9.9 \tablenotemark{b} & 3.52 \\
     Ni  I &  6767.78 &  -2.170 &  1.826  &     --  &  --    &    7.0 \tablenotemark{b} & 3.40 \\
     Cu  I &  5105.55 &  -1.520 &  1.390  &     syn  &  $<$0.94  &   syn  & $<$1.31 & \\
     Zn  I &  4722.15 &  -0.390 &  4.030  &     7.8 & 2.00 &   20.5 & 2.41 \\
     Zn  I &  4810.53 &  -0.170 &  4.078  &    11.6 & 2.02 &   27.5 \tablenotemark{b} & 2.43 \\
     Sr II &  4077.72 &   0.150 &  0.000  &   121.0 & -0.09 &   86.5 \tablenotemark{b} & -0.87 \\
     Sr II &  4215.54 &  -0.180 &  0.000  &   117 \tablenotemark{a}  & 0.12     &   86.0 & -0.62 \\
     Ba II &  4554.04 &   0.170 &  0.000  &    20.2  & -1.81 &   20.2  & -2.74 \\
     Ba II &  4934.10 &  -0.150 &  0.000  &    46.0  & -1.77 &   13.3  & -2.68 \\
     Eu II &  3819.67 &   0.510 &  0.000  &    syn    & -2.98   & --     & -- \\
     Eu II &  4129.72 &   0.220 &  0.000  &    syn    & -2.78   & syn    & $<-$2.66 \\
     Eu II &  4205.05 &   0.210 &  0.000  &    syn    & -3.00   & --     & -- \\
     
\enddata
\tablenotetext{a}{Taken from McWilliam et al (1995).}
\tablenotetext{b}{Measured for the spectrum obtained in 2004.}

\end{deluxetable}

\begin{deluxetable}{lcccc}
\tablewidth{0pt}
\tablecaption{ADOPTED ATMOSPHERIC PARAMETERS \label{tab:param}}
\tablehead{
Star    & {\teff}(K) &  {\logg} & [Fe/H] & {\vt}(km s$^{-1}$) 
}
\startdata
{\BS} & 4760 & 1.2 & $-3.12$ & 1.4 \\ 
{\HD} & 4810 & 1.8 & $-2.89$ & 1.6 
\enddata
\end{deluxetable}

\begin{deluxetable}{cccccccccccc}
\tablewidth{0pt}
\tablecaption{ABUNDANCE RESULTS\label{tab:res}}
\tablehead{
Element & Species  &  Solar Abundance\tablenotemark{a} & \multicolumn{4}{c}{\HD}           & & \multicolumn{4}{c}{\BS} \\
\cline{4-7}\cline{9-12}
        &          & {\loge}          & {\loge} & [X/Fe] & $N$ & $\sigma$ & &  {\loge} & [X/Fe] & $N$ & $\sigma$ 
}
\startdata
C       & CH \tablenotemark{b} & 8.43 & 5.78 &   +0.20 & \nodata & \nodata    & & 5.37   & +0.12 & \nodata & \nodata \\ 
Na      & \ion{Na}{1} &   6.24    & 3.97    &   +0.58  & 2    & \nodata   & & 3.24   &   +0.18 & 2  & \nodata \\ 
Mg      & \ion{Mg}{1} &   7.60    & 5.17    &   +0.42  & 7    &  0.13     & & 4.66   &   +0.24 & 7       &  0.21\\ 
Al      & \ion{Al}{1} &   6.45    & 3.09    & $-$0.51  & 1    & \nodata & & 2.66   & $-$0.61 & 1  & \nodata \\ 
Si      & \ion{Si}{1} &   7.51    & 5.08    &   +0.42  & 1    & \nodata & & 4.23   & $-$0.10 & 1  & \nodata \\ 
Ca      & \ion{Ca}{1} &   6.34    & 3.94    &   +0.45  & 8    &  0.18     & & 3.26 &   +0.10 & 10 &  0.14\\ 
Sc      & \ion{Sc}{2} &   3.15    & 0.42    &   +0.12  & 3    &  0.03     & & $-$0.26 &$-$0.23 & 3 &  0.08\\ 
Ti      & \ion{Ti}{1} &   4.95    & 2.36    &   +0.26  & 10   &  0.08     & & 2.18 &   +0.41 & 11  &  0.13\\ 
Ti      & \ion{Ti}{2} &   4.95    & 2.59    &   +0.49  & 20   &  0.13     & & 2.26 &   +0.49 & 20  &  0.14\\ 
V       & \ion{V}{1}  &   3.93    & 1.08    &  ~~0.00  & 1    & \nodata   & & 0.59 & $-$0.16 & 1   & \nodata \\ 
V       & \ion{V}{2}  &   3.93    & 1.30    &   +0.22  & 2    & \nodata   & & 1.15 &   +0.40 & 1   & \nodata \\ 
Cr      & \ion{Cr}{1} &   5.64    & 2.67    & $-$0.12  & 5    &  0.29     & & 2.62   & +0.16 & 6   &  0.20 \\ 
Cr      & \ion{Cr}{2} &   5.64    & 2.94    &   +0.15  & 2    & \nodata   & & 2.52   & +0.06 & 2   &  \nodata \\ 
Mn      & \ion{Mn}{1} &   5.43    & 1.83    & $-$0.75  & 6    &  0.27     & & 2.07   & $-$0.18 & 6 &  0.33\\ 
Fe      & \ion{Fe}{1} &   7.50    & 4.64    & $-$2.86 \tablenotemark{c} & 82 & 0.12  & & 4.38 & $-$3.12 \tablenotemark{c}& 84 & 0.22 \\
Fe      & \ion{Fe}{2} &   7.50    & 4.66    & $-$2.84 \tablenotemark{c} & 13 & 0.11  & & 4.26 & $-$3.24 \tablenotemark{c}& 10 &  0.17\\ 
Co      & \ion{Co}{1} &   4.99    & 2.29    &   +0.15  & 7    &  0.13     & & 2.13   &   +0.32 & 5   &  0.20\\ 
Ni      & \ion{Ni}{1} &   6.22    & 3.35    & $-$0.02  & 3    &  0.03     & & 3.48   &   +0.44 & 5  &  0.06 \\ 
Cu      & \ion{Cu}{1} &   4.19    & $<$0.94    & $<-$0.40  & 1    & \nodata   & & $<$1.31   & $<$+0.30 & 1  &  \nodata \\ 
Zn      & \ion{Zn}{1} &   4.56    & 2.01    &   +0.30  & 2    & \nodata   & & 2.42   &   +1.04 & 2  &  \nodata \\ 
Sr      & \ion{Sr}{2} &   2.87    & 0.02    &  ~~0.00  & 2    & \nodata   & &$-$0.75 & $-$0.44 & 2  & \nodata \\ 
Ba      & \ion{Ba}{2} &   2.18    &$-$1.79  & $-$1.12  & 2    & \nodata   & &$-$2.71 & $-$1.71 & 2       &  \nodata\\
Eu      & \ion{Eu}{2} &   0.52    &$-$2.92  & $-$0.59  & 3    & 0.12      & &$<-$2.66   & $<$+0.00 & 1  &  \nodata \\ 

\enddata
\tablenotetext{a}{\citet{asplund09}}
\tablenotetext{b}{Derived from the CH 4320~{\AA} band}
\tablenotetext{c}{[Fe/H] values}
\end{deluxetable}

\clearpage

\begin{figure} 
\includegraphics[angle=-90,width=16cm]{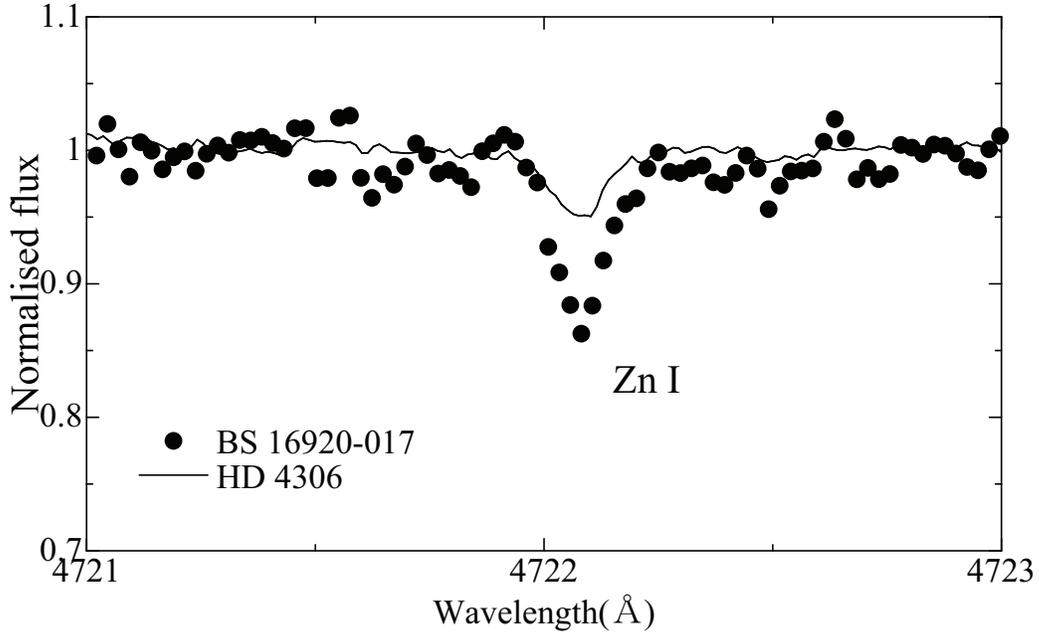} 
\includegraphics[angle=-90,width=16cm]{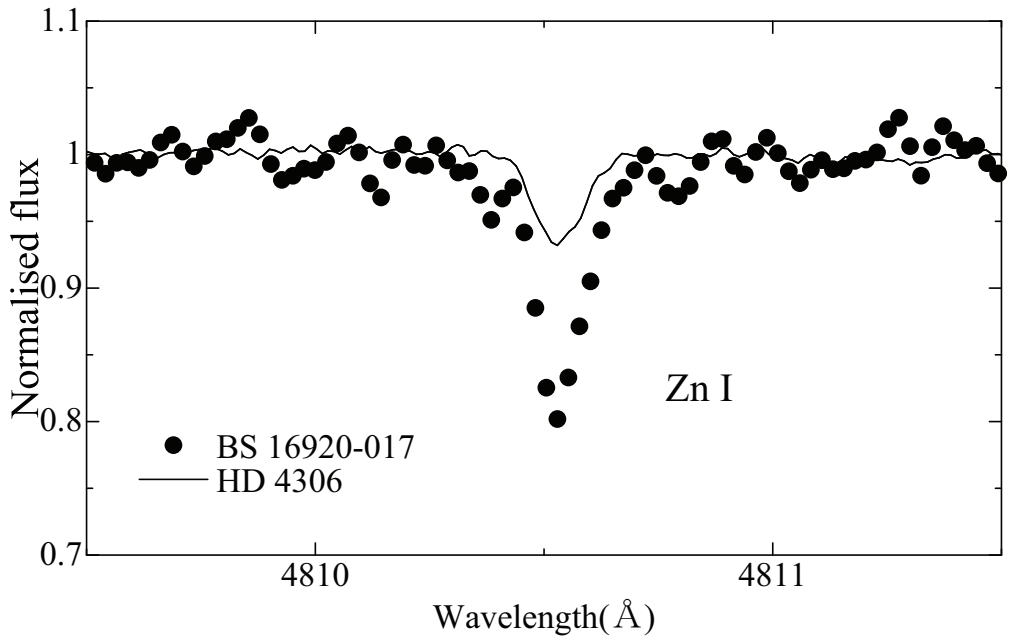} 
\caption[]{Comparison of the spectra of the Zn $\lambda$4722~{\AA} line in
  {\BS} (filled circles) and {\HD} (solid line). Although the metallicity of
  {\HD} is higher than {\BS}, the Zn line is clearly stronger in {\BS}.}
\label{fig:spec3}
\end{figure}

\begin{figure} 
\includegraphics[angle=-90,width=16cm]{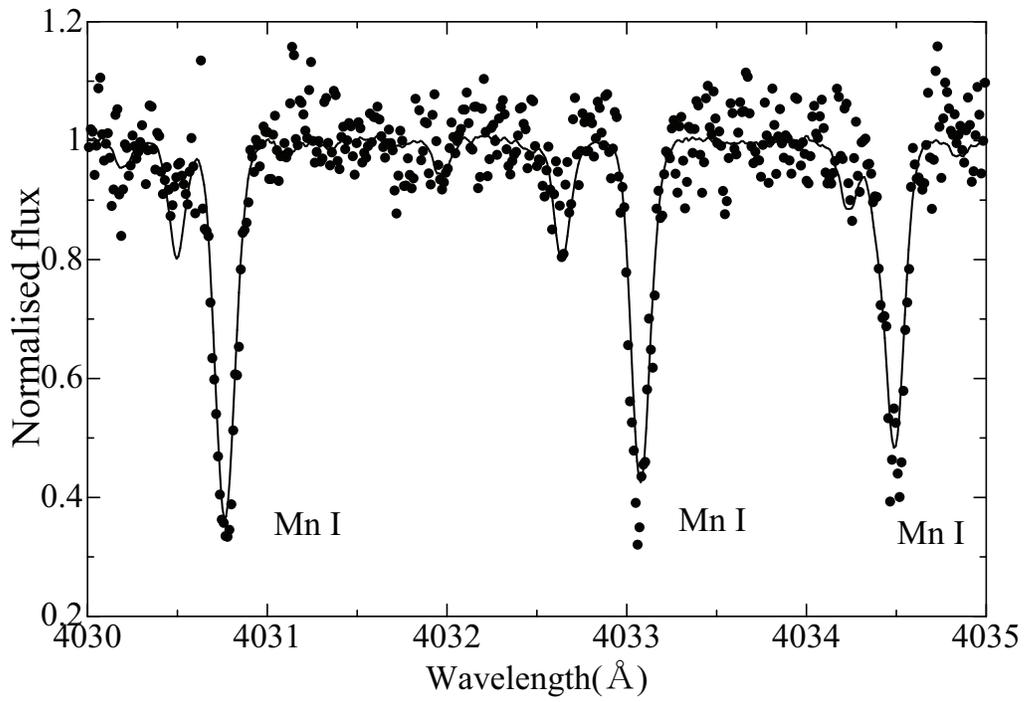} 
\caption[]{Comparison of the spectra of the Mn lines in
  {\BS} (filled circles) and {\HD} (solid line). Although the metallicity of
  {\HD} is higher than {\BS}, two of the three Mn lines are clearly stronger in {\BS}.}
\label{fig:mn} 
\end{figure}

\begin{figure} 
\includegraphics[angle=-90,width=16cm]{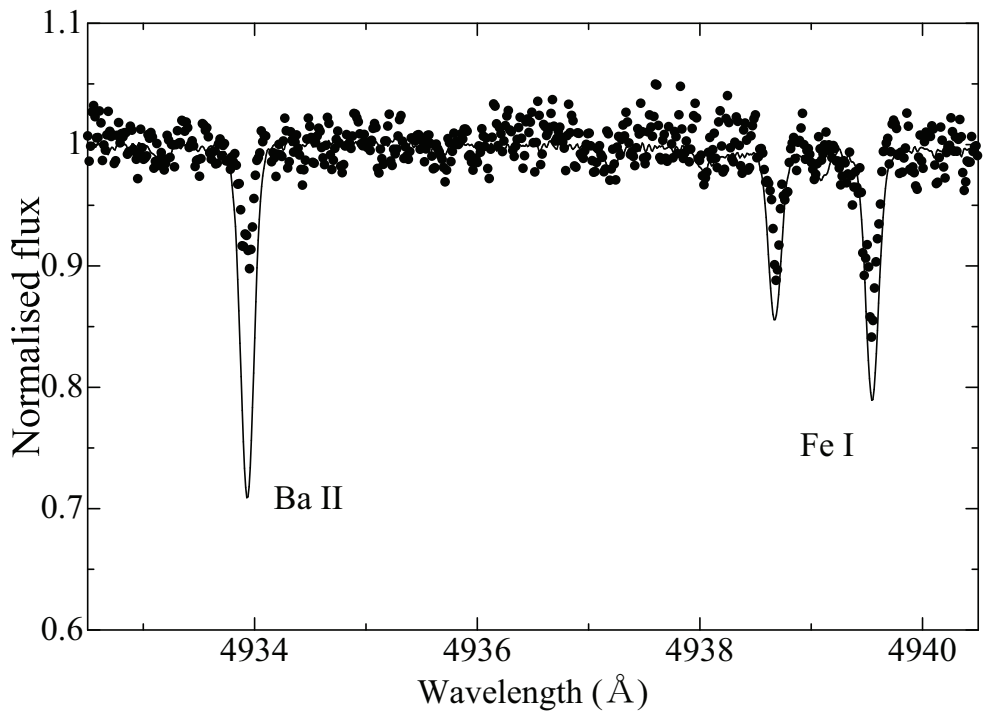} 
\caption[]{Comparison of the spectra of the Ba and Fe lines in
  {\BS} (filled circles) and {\HD} (solid line). The Ba line of {\BS} is
  significantly weaker than that of {\HD}.}
\label{fig:ba} 
\end{figure}

\begin{figure} 
\includegraphics[angle=-90,width=16cm]{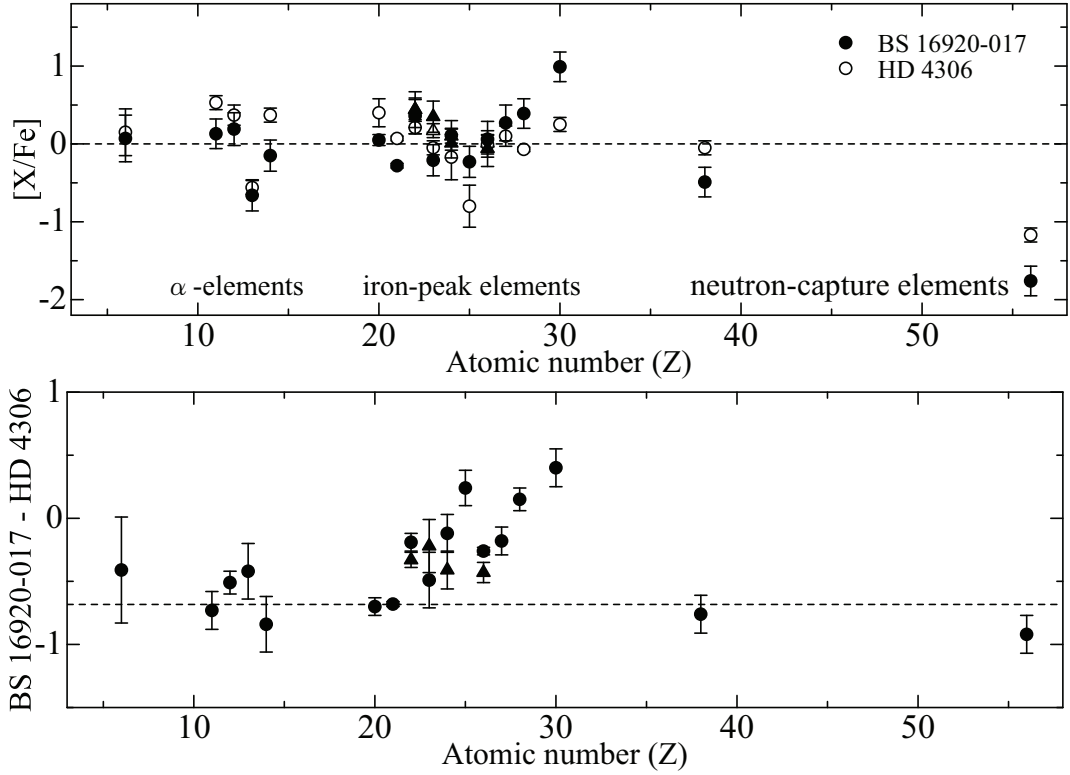} 
\caption[]{{\it Upper panel}: Comparisons of the abundance ratios, [X/Fe], as a
  function of the atomic number ($Z$), for {\BS} (filled circles and triangles) and
  {\HD} (open circles and triangles). The circles indicate abundances derived from 
  neutral species and triangles indicate those from ionized species (for Ti, V,
  Cr, and Fe). The dashed line is a reference at [X/Fe] = 0.0. 
  {\it Lower panel}: Abundance differences, on the logarithmic scale, 
  between the two stars. The dashed line indicates the average of the 
  values of the $\alpha$ elements (Mg, Si, and Ca).

}

\label{fig:abund}

\end{figure}

\end{document}